\documentclass[conference]{IEEEtran}
\IEEEoverridecommandlockouts
\usepackage[left=1.62cm,right=1.62cm,top=1.9cm]{geometry}
\usepackage{amsmath,amssymb,amsfonts}
\usepackage{algorithmic}
\usepackage{graphicx}
\usepackage{textcomp}
\usepackage{xcolor}
\usepackage{gnuplottex}
\usepackage{float}
\usepackage{svg}

\def\BibTeX{{\rm B\kern-.05em{\sc i\kern-.025em b}\kern-.08em
    T\kern-.1667em\lower.7ex\hbox{E}\kern-.125emX}}

\usepackage[shortcuts,acronyms]{glossaries}
\usepackage[backend=biber, sorting=none]{biblatex}
\newtheorem{lemma}{Lemma}
\newtheorem{remark}{Remark}

\newacronym{RCU}{RCU}{Robot-Compute Unit}
\newacronym{OJDR}{OJDR}{Optimal Joint Detection Receiver}
\newacronym{OSSR}{OSSR}{Optimal Single Shot Receiver}
\newacronym{OEAR}{OEAR}{Optimal Entanglement-Assisted Receiver}
\newacronym{M2M}{M2M}{Machine to Machine}
\newacronym{M2C}{M2C}{Machine to Cloud}
\newacronym{CDMA}{CDMA}{Code Division Multiple Access}
\newacronym{QIP}{QIP}{Quantum Information Processing}
\newacronym{OFDM}{OFDM}{Orthogonal Frequency Division Multiple Access}
\newacronym{mMIMO}{mMIMO}{massive Multiple-Input Multiple-Output}

\title{
Effects of Quantum Communication in Large-Scale Networks at Minimum Latency
}
\author{Sekavčnik, Simon and Nötzel, Janis

\thanks{\copyright 2022 IEEE. Personal use of this material is permitted. Permission from IEEE must be obtained for all other uses, in any current or future media, including reprinting/republishing this material for advertising or promotional purposes, creating new collective works, for resale or redistribution to servers or lists, or reuse of any copyrighted component of this work in other works.}
}

\date{March 2022}
\begin{document}
\maketitle

\begin{abstract}
    Quantum communication technology offers several advanced strategies. However, their practical use is often times not yet well understood. In this work we therefore analyze the concept of a futuristic large-scale robotic factory, where each robot has a computing unit associated to it. The computing unit assists the robot with large computational tasks that have to be performed in real-time. Each robot moves randomly in a vicinity of its computing unit,  and in addition both the robot and the unit can change location. To minimize latency, the connection is assumed as optical wireless. Due to the mobility, a permanent optimal assignment of frequency bands is assumed to increase communication latency and is therefore ruled out. Under such assumptions, we compare the different capacity scaling of different types of such architectures, where the one is built utilizing quantum communication techniques, and the other based on conventional design methods.
\end{abstract}

\section{Introduction}

\IEEEPARstart{R}{obots} in a factory need low latency communication with their respective computing units, therefore communication over the air is preferred over the fiber communication. Coordination and synchronisation, although desirable in a multi-access scenario, is time-consuming and therefore adds to latency, induced from the communication between parties that is necessary to achieve a coordinated strategy \cite{willems}. In a situation with multiple \ac{RCU}s an interference between the pairs will arise if the communication is not synchronised.

This concept can be traced back several years to works such as \cite{rcuConcept}, where the inherent computational limitations of robotic units, inherited from the robot size, shape, power supply, motion mode, and working environment are discussed and the topic of upgrading or even just adapting robotic computing performance after the robot is built, is brought up. As a solution to such problems, computational offloading is discussed. The work \cite{rcuConcept} suggests a separation of the communication into an \ac{M2M} layer where robots communicate amongst themselves and an \ac{M2C} layer where they communicate with the cloud.

As the focus of the present work is to clarify a hypothetical use of quantum technologies by showing a separation of performance under a clean cut futuristic use case, we consider a special case where no communication takes place on the \ac{M2M} layer (only offloading of computational problems from robot to computing unit), and where the communication on the \ac{M2C} layer is divided into two planes. The first plane is realized through a  free space optical- or LiFi link, and is assumed to induce the minimum possible latency, but suffers from noise. The second plane is realized for example via optical fiber. It is assumed to deliver low noise connectivity, but does not have to obey any latency constraint.  
Based on such a design, we can show the different properties of three types of communication links in comparison. In all three cases, our main focus is on the low-latency wireless \ac{M2C} communication part only. In all three cases, the physical properties of the links under study are characterized by loss $\tau$, transmit power $n$, baud-rate (number of pulses per second) $b$ and (thermal) noise $\nu$.

The first link uses established communication techniques. Given a specified amount of available spectrum, loss and power, it and operates at the Shannon limit. We call such a system an \ac{OSSR}.

The second link uses a \ac{OJDR}. This system is therefore assumed to operate at the Holevo limit. Following \cite{ntzel2022operating}, such a link can be expected to outperform the first link in situations where it utilizes a large amount of spectral bandwidth.

The third link uses an \ac{OEAR}. It uses the second plane of the \ac{M2C} link to generate entanglement, which is then utilized as a way to boost transmission capacity on the wireless \ac{M2C} link.

To avoid any delays arising from coordination, we study two extreme approaches to assigning spectral bandwidth to the \ac{RCU}s: In the first one we use \ac{OFDM} to assign an individual slice of the available spectrum to each participant. In the second one we base our analysis on \ac{CDMA}, so that every \ac{RCU} can utilize the entire available spectrum. In both cases, we do not discuss details of an implementation.

Studying these three systems in comparison in the envisioned scenario is interesting for the following reasons: \ac{QIP} has, for communication systems, so far pointed out the existence of infinite-fold gains \cite{guha2020infinite} - however, most of these arise in parameter regimes $(b,n,\tau,\nu)$ which are never realized in practice. However, it is obvious from the literature \cite{highBaudrateComms} that e.g. baud-rates in the established optical fiber links have increased steadily over time. Therefore, our analysis aims to point out how \ac{QIP} can start to play a role in future networks \emph{given certain trends}.

To explain the basics of the approach let us clarify the differences between the three different links. In all cases where the product $\tau n/b$ is small enough (for example below one), the \ac{OJDR} will eventually display a logarithmic advantage over the \ac{OSSR} \cite{ntzel2022operating} that scales with $\tau n/b$. This clarifies how a trend towards increasing baud-rates can establish \ac{QIP} as the technology of choice - if \ac{CDMA} outperforms \ac{OFDM} in our selected use case. 

Further, it has been shown \cite{guha2020infinite} that in cases where $\nu\gg1$ and $n/b\ll1$, the \ac{OEAR} outperforms even the \ac{OJDR} by a factor which again scales as $-\log(n/b)$. While current commercial systems satisfy $n/b\gg1$, the condition $\nu\gg1$ still fuels some hope that a design based on \ac{CDMA} and the \ac{OEAR} could eventually be the superior choice. For such a system, we show some hypothetical advantage utilizing the concept of \ac{mMIMO}.



\section{Setup}

Each \ac{RCU} is characterized by a loss parameter $\tau$, its transmit power of $n$ photons per second, the baud-rate $b$ (or symbol rate) and the amount of (thermal) noise $\nu$ affecting the link. 
The total available spectrum in the factory is $B$, and since $b$ dictates not only the number of symbols per second but also the spectral width of the signal it must hold $b\leq B$. All parameters are assumed to be equal for each \ac{RCU}. The $K^2$ units are placed on a $K\times K$ grid with vertical and horizontal distance between neighbouring nodes equal to $d$. We consider two types of loss: In the pure loss model the interference power scales as 
    \begin{equation}
        P_r^L = \tau_{\ac{RCU}}\frac{n}{r^2}
    \end{equation}
     In the two-ray loss model \cite{groundEffect}, we consider shadowing and scattering effects, which lead to an interference power of 
    \begin{equation}
        P_r^S = \tau_{\ac{RCU}}\frac{n}{r^4}
    \end{equation}
    between \ac{RCU}s separated by a distance of $r$. Interference is only relevant when \ac{CDMA} is used, it equals zero otherwise.
\begin{figure}[H]
    \includegraphics[width=1\columnwidth]{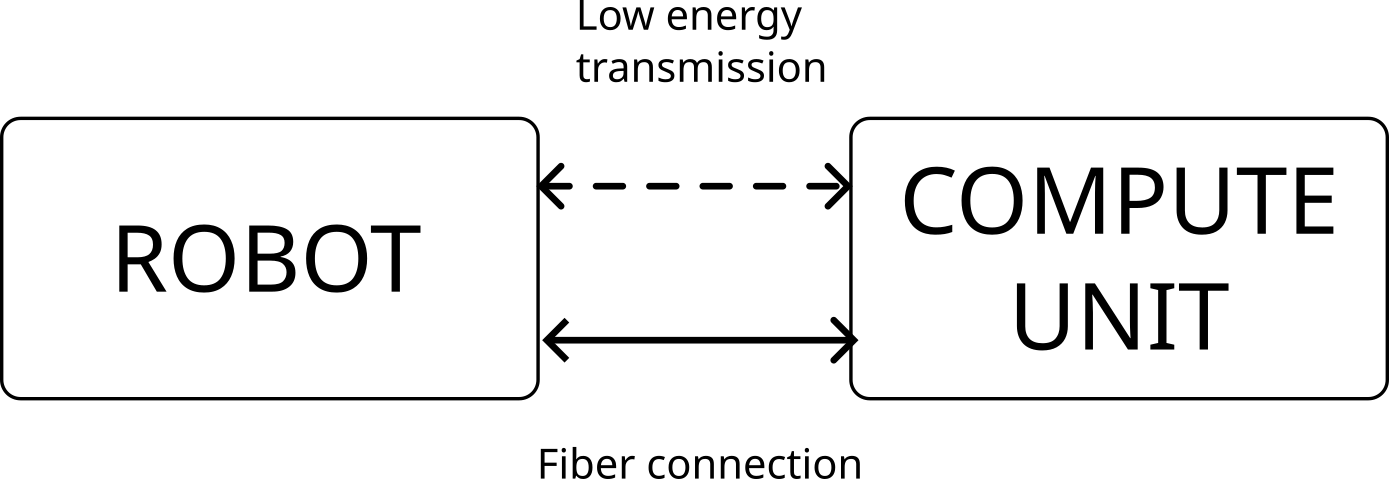}
    \caption{The robot arm might carry a small communication unit, which to avoid additional costs, resulting in low energy transmission. This situation necessarily results in a communication link with a high noise and low transmission energy. In the case where \ac{OEAR} is used, we may assume an additional fiber connection, which has a higher latency but provides enough capacity to fill entanglement buffers at the end of the \ac{RCU} pair.}
\end{figure}
    For numerical results we assume the equipment is operating in the $C$-band around $1550nm$ where standard equipment is cheaply available and set $B=10^{12}s^{-1}$. The energy per photon is then $\approx 1.3\cdot 10^{-19}J$. Correspondingly, a source radiating $1W$ of power emits $\approx 7.8\cdot10^{18}$ photons per second. We calculate the transmittivity based on the De Friis model in equation \eqref{eq:tau} as  $\tau_{\ac{RCU}}=1.5 \cdot 10^{-6}$ and $\tau=10^{-7}$, for which we chose the gain of both antennas to equal $10^4$ ($40dB$) and set the distance between \ac{RCU} nodes $10m$ and the distance between \ac{RCU}s to $20m$. Transmission power equals $1mW$, corresponding to $n=7.8 \cdot 10^{15}$, leading to $\tau n/b \approx 10^{-3}\ll1$ but $n/b\approx10^{4}\gg1$.
    

    If \ac{CDMA} is used then each \ac{RCU} experiences interference power $\nu/b$ where $\nu$ is equal to 
\begin{equation}\label{eqn:interference-power}
    P_I^X := \sum_{(x,y)\in \mathcal{N}} P_r^X\qquad \mathrm{(\ where\ }X\in\{L,S\}\ )
\end{equation}
    where $\mathcal{N}$ is the set of coordinates of all \ac{RCU}s, except the one where the noise is being measured. Thermal noise is accounted for by adding $\nu_{TH}=10^9/b$ noise photons for the entire C-band, a number which is derived from \cite[Eq. (1)]{antennaNoiseFormula}. This corresponds to a situation where the factory is at a temperature of $300K$ without sunlight.

Each \ac{RCU} can either use the \ac{OSSR}, the \ac{OJDR} or the \ac{OEAR}. Given parameters $b, n, \tau, \nu$, the corresponding capacities \cite{holevo2012quantum,shannon1948mathematical,bennett1999entanglement} are:
\begin{align}
    C_{J}(b,\tau,n,\nu) &= \Big(g(\frac{n+\nu}{b})-g(\frac{\nu}{b})\Big) \cdot b \label{eq:C_J}\\
    C_{S}(b,\tau,n,\nu) &= \log{\Big(1+\frac{\tau\cdot n}{b+\nu} \Big)}\cdot b \label{eq:C_S}\\
    C_{E}(b,\tau,n,\nu) &= \sum_{x=0,1}\left(g(\tfrac{\tau\cdot n+x\cdot\nu}{b})- g(d_x(\tau,\tfrac{n}{b}, \tfrac{\nu}{b}))\right)b \label{eq:C_E}
\end{align}
where $g(x)=(x+1)\log(x+1) - x\log x$ and 
\begin{align}
    d_{x}(\tau, n, \nu) &= (d(\tau, n, \nu) - 1 + (-1)^x ( (\tau-1) n + \nu ))/2\\
    d(\tau, n, \nu)&=\sqrt{(( 1 + \tau ) n + \nu +1 )^2 - 4\tau n( n + 1 ) }.
\end{align}
The issue of entanglement generation, distribution and storage is omitted. In the model entanglement is always available to the \ac{OEAR} based link.

\section{Results}
\begin{figure}[H]
    \centering
    \input{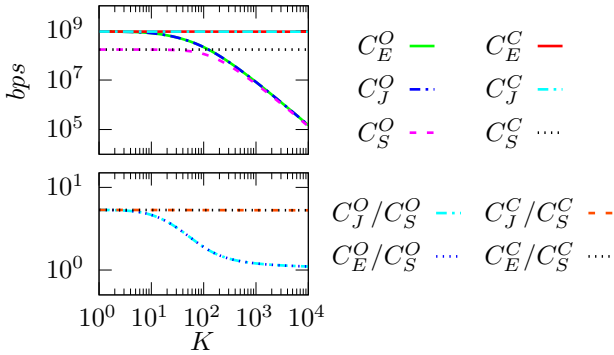}
    \caption{One link capacity scaling within the increasing size of the factory grid, where grid distance is set to $d=10m$ and communicating link distance is set to $r=10m$. We observe no capacity advantage when comparing \ac{OEAR} with \ac{OJDR}. We do, however, observe advantage of any quantum technique over the classical approach. In case of \ac{CDMA} the advantage is more pronounced.}
    \label{fig:results_pure_loss}
\end{figure}

For both interference models we observe a roughly tenfold increase in link capacity (see Figure \ref{fig:cdma_free_space}) when the \ac{OJDR} is used instead of the \ac{OSSR} and both use \ac{CDMA}. For the pure loss scenario ($X=L$) the interference power scales as $P_I^L\approx n\cdot \log K$, which leads to a slight decrease of the observed advantage towards a factor of $5$ when $K=10^4$. For the two-ray loss model, $P_I^S$ converges to a constant when $K$ grows and the advantage is constant at large $K$. 
For the pure loss case our numerical results indicate that a hierarchy can be established according to which \ac{CDMA} has a superior scaling over \ac{OFDM}, when equal link technologies are compared (see figure \ref{fig:results_pure_loss}). 
As our analysis is matched to the power budgets and baud-rates of current equipment, the requirement $n/b\gg1$ for the \ac{OEAR} to show its superiority over the \ac{OJDR}, is not satisfied. 
We conclude the \ac{OEAR} can provide advantage with a different link design (\ac{mMIMO}, see Section \ref{sec:massiveMimo}, Figure \ref{fig:results_mMimo}), where instead of one antenna an antenna array is used under a global power budget and without coordination amongst antennas. For the individual antennas, such approach can lead to a situation where the requirement $n/b\ll1$ is fulfilled for the individual antennas. In addition, we point out the possibility for utilizing the technology for wireless on-chip networks \cite{yu2014architecture} where lower transmission powers might be used.

\begin{figure}[H]
    \centering
    \input{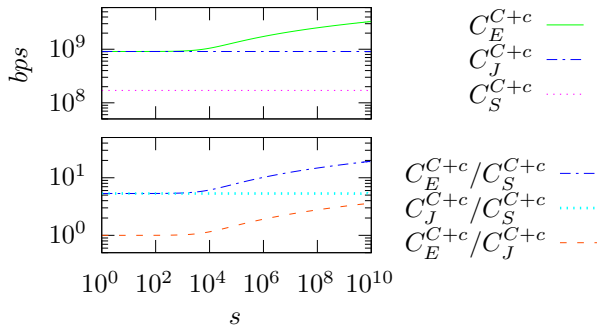}
    \caption{Capacity of one link improves with \ac{mMIMO} approach with \ac{OEAR} in the case of \ac{CDMA}. The entanglement assisted capacity of one \ac{mMIMO} link, although suffering with small antenna arrays, scales with the antenna array size, whereas \ac{OJDR} and \ac{OSSR} capacity stagnate. The factory size is fixed to $K=100$.}
    \label{fig:results_mMimo}
\end{figure}


\section{Methods}

\subsection{Interference noise scaling with pure loss factory model}
If we assume an \ac{RCU} unit is in the center of a rectangular $K \times K$ grid with distance $d$ between the grid points, then in the pure loss case the total averaged interference power that has to be tolerated by that unit can be calculated as:
 \begin{align} \label{eq:fs_loss_interference}
 \mathcal{I}(K) :&= \frac{4}{d^2} \sum_{i=1}^{K} \Big( \frac{1}{i^2} + \sum_{k+l=i}\frac{1}{k^2+l^2}\Big)\\
 &= \frac{4}{d^2} \Big( H_{K,2} + \sum_{i=1}^{K} \sum_{k+l=i} \frac{1}{k^2+l^2}\Big)
 \end{align}
 
 where all the sum are over natural numbers, $i,k,l \in \mathbb{N} $ and $\lim_{K\to \infty}=\pi^2/6$. The sum can be upper and lower bounded by integrals as $\frac{C}{d^2}(\log(K)+C_X)$ for suitable constants $C, C_L, C_U$, through elementary calculus. Thus neglecting constant terms we get an asymptotic scaling of $I(K)$ as $|I(K)-\frac{C}{d^2}\log K| \leq  \max\{C_U, C_L\}$. Given an \ac{RCU} factory of size  $K\times K$ only one \ac{RCU} will be exactly in the center. If the factory is large enough though, almost all \ac{RCU}s will be in the center of a \ac{RCU} grid of size at least $K^\alpha \times K^\alpha$, where $\alpha \in (0,1)$ due to the following lemma:
 
 \begin{lemma}
    Let $L,K\in \mathbb{N}$ satisfy $L<K$. The number of elements in a boundary of size $L$ is $2\cdot L \cdot (K-L)$. The probability $\mathbb{P}(L,K)$ that a randomly selected $L\times L$ sub-grid $S$ of a $K \times K$ grid $G$ is contained entirely within $G$ is given by:
 \begin{equation}
     \mathbb{P} (L,K) = \Big(\frac{K-L}{l}\Big)^2 =  \Big(\frac{K}{L}-1\Big)^2
 \end{equation}
 \end{lemma}
 \begin{remark}
     Due to its obvious simplicity, above Lemma is stated without proof.
 \end{remark}
\begin{remark}
    Thus for every functional dependence $K \to L(K)$  for which $\lim_{K \to \infty}K/L(K)=0$, for example $L(K)=K^\alpha$ with $\alpha \in (0,1)$, the probability of having interference power at least $\alpha \cdot C \cdot C \cdot \frac{P}{d^2}\log(K)$ approaches 1 as $K \to \infty$. Thus the logarithmic scaling of interference power with the factory size is accurate with probability approaching $1$ as $K\to\infty$.
\end{remark}

\subsection{Calculation of transmittivity\label{subsec:calculationOfTransmittivity}}
To calculate the loss we use the formula

\begin{align}\label{eq:tau}
    \tau = \frac{P_r}{P_t} = G_t \cdot G_r \Big( \frac{\lambda}{4 \pi r} \Big)^2
\end{align}

with $G_t, G_r$ being antenna gains at transmitter and receiver, and $\lambda$ being the carrier wavelength \cite{friis1946note}. For calculations we use $G_t=G_r=10^{4}$ and $\lambda=1550nm$, so that 
\begin{align}\label{eqn:specific-tau}
    \tau(r)=1.5 \cdot 10^{-6}r^{-2}
\end{align}
    
\subsection{Available Resources and Environment}
    The baud-rate $b$ is limited by the total spectral bandwidth $B$ available in the network as $b\leq B$. The noise $\nu$ depends on the parameters $K$ and $d$ which model the size and the density of the network, as well as on the transmit power $P$ of the individual units. In addition, each unit experiences an amount ${\nu_{TH}=10^{-9}/b}$ of noise per second radiated from the environment, a number which is derived from \cite[Eq. 1]{antennaNoiseFormula}.

\subsection{Resource Allocation}
    In this work, our main focus is on low-latency communication. At any given moment in time, the various RCUs form an interference channel \cite{negro2010mimo}. For such a channel it is known \cite{willems} that communication schemes exist which yield communication rates strictly above those that are obtained by treating, for every RCU $k$, its $K-1$ neighbours as interfering noise. However, setting up the respective code-books is inevitable a time-consuming task. Therefore coordination between \ac{RCU}s is set to zero in this initial work. A study of trade-offs between coordination and low-latency data transmission is left to future work. Instead, we concentrate first on the comparison of three communication methods (the \ac{OEAR}, the \ac{OJDR} and the \ac{OSSR}) which all operate in a simplistic fashion, by treating all neighbouring RCUs as noise. The three methods vary in the degree to which they use, on the individual links, quantum methods to enhance the capacity. The question of finding the actual sweet spot for low-latency communication involving also coordination methods is left open.
    With regards to the allocation of spectrum, we use two different non-coordinated methods: In the first method (\ac{OFDM}), every \ac{RCU} gets a channel of spectral width equal to $B/K^2$ (by design, $K^2$ is the number of \ac{RCU}s in the network). When \ac{OFDM} is used, the channels of different \ac{RCU}s do not interfere. In the second method (\ac{CDMA}), every \ac{RCU} gets to transmit on the entire spectrum with a spectral width of $B$. In this second method, \ac{RCU}s cause interference with power \eqref{eqn:interference-power} to each other.

\subsection{Capacity scaling with pure loss model}

\subsubsection{Orthogonal Frequency Division Multiplexing}

With \ac{OFDM} the $(b, \tau, n, \nu)$ parameters are chosen by the following procedure.
Individual baud-rates are determined by the fraction of the total bandwidth of the factory $b=B/K^2$, where $K$ denotes the size of the factory with $K^2$ communicating \ac{RCU} pairs. Each \ac{RCU} is emitting a signal with power $n\approx 7.8 \cdot 10^{15}$, corresponding to $1mW$. 

The basic assumption in the \ac{OFDM} type transmission is that the communication bands between any two \ac{RCU}s are non overlapping. As a result of allowing only very narrow bands for the \ac{RCU}s to communicate, no interference is experienced between links of different \ac{RCU} pairs. In effect the only noise that is experienced by the communicating \ac{RCU}s is the thermal noise $\nu_{TH}= 10^{-9}/b$.
 
The noise per pulse experienced by the \ac{OFDM} receivers, is invariant to the size of the factory, the reason for this is, that as bands get narrower, also the baud-rates get smaller, thus per pulse noise is constant. 

\begin{figure}[H]
\centering
\input{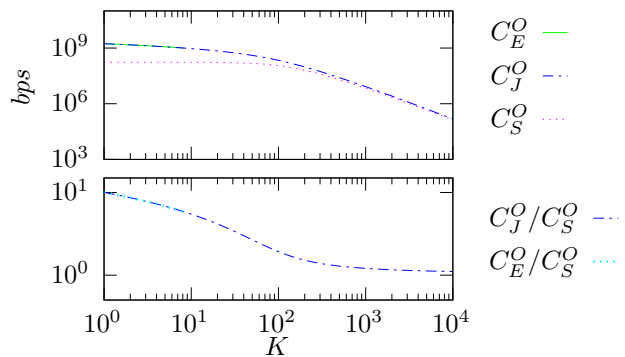}
\caption{Scaling of \ac{RCU} communication capacity in a factory which is using \ac{OFDM}. The grid distance is set to $d=10m$ and the communicating link distance is set to $r=10m$. The noise per pulse experienced by a single receiver is $\nu_{TH}\cdot b/b= 10^{-9}$.}
\label{fig:ofdm_free_space}
\end{figure}

The transmission power ($n \approx 7.8 \cdot 10^{15}$ photons per second) stays constant with growing number of the \ac{RCU} pairs as the bandwidth and consequently the baud-rates get lower. For each single \ac{RCU}, the ratio $\tau n/b \approx 10^{-4}\cdot K^2$ is therefore not favorable for \ac{QIP} methods as soon as $K>100$. The scaling of \ac{OFDM} can be observed in Figure \ref{fig:ofdm_free_space}, where the difference between the capacity of the \ac{OJDR} (and \ac{OEAR}) and the \ac{OSSR} becomes minute for $K>100$. 

\subsubsection{Code Division Multiple Access}

With \ac{CDMA} the \ac{RCU} pairs make the opposite trade off. They all communicate utilizing the whole bandwidth, $b=B$. In this case, due to the lack of coordination between the \ac{RCU}s, the communication signals from neighbouring \ac{RCU}s are detected as noise, which is calculated using equation \eqref{eq:fs_loss_interference} and added to the background noise $\nu_{TH}$ which then evaluates to $\nu_{TH}=10^{-3}$.

\begin{figure}[H]
    \centering
    \input{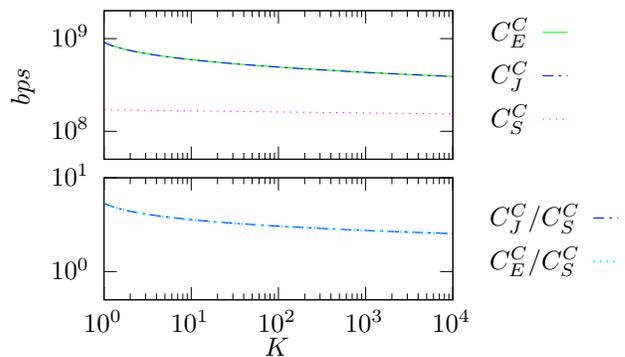}
    \caption{Scaling of \ac{RCU} communication capacity in a factory which is using \ac{CDMA}. The grid distance is set to $d=10m$ and the communicating link distance is set to $r=10m$. The advantage offered by the \ac{QIP} is clearly visible in the chosen parameter regime, but slowly decreasing with rising factory size. There is no \ac{OEAR} advantage over the \ac{OJDR} in this parameter regime.}
    \label{fig:cdma_free_space}
\end{figure}

In the case of \ac{CDMA} the ratio $\tau n/b \approx 1.2 \cdot 10^{-4}$ is way below one and is invariant to the factory size parameter. The corresponding advantage can be  observed in Figure \ref{fig:cdma_free_space}. 

There is no advantage of the \ac{OEAR} over \ac{OJDR}, because both the ratio $n/b \approx 7.8 \cdot 10^3$ and the noise $\nu<1$ are in an unfavorable region.

\subsection{Capacity scaling with two ray loss model}

In a factory with double path fading signal propagation, we can study communication capacity scaling as the factory gets denser. In this case, we consider as interference power the limiting distribution $P^L_{I,\infty}$ when $K\to\infty$, which equals
\begin{align}
    P^L_{I,\infty}&= 4 \cdot \tau_{RCU} \frac{n}{d^4} X,\label{eqn:interference-power-in-two-ray-loss-model}
\end{align}
where $X\approx0.42$. In this scenario only the \ac{CDMA} case is considered, as the capacity in the \ac{OFDM} case would vanish due to infinitely narrow bandwidths.

\begin{figure}[H]
    \centering
    \input{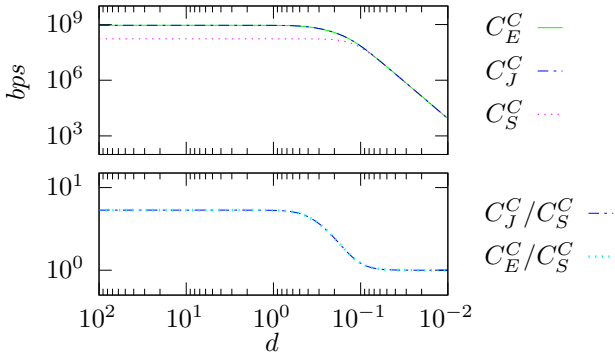}
    \caption{\ac{CDMA} link with the distance $r=10m$ capacity scaling in relation to the density (grid distance $d$ ) with of one link inside of infinitely big factory with two-ray loss model.}
    \label{fig:cdma_2_ray}
\end{figure}

With the two ray loss model all of the capacities scale  (see Figure \ref{fig:cdma_2_ray}) anti-proportional with the rising grid density $1/d$. As it does not depend on the value of $d$, the estimate $\tau n/b \ll 1$ is satisfied again for the whole plotted region. When the factory gets more dense and the interference power \eqref{eqn:interference-power-in-two-ray-loss-model} increases, the advantage of \ac{QIP} techniques decreases.

\section{Scaling with link massive MIMO\label{sec:massiveMimo}}

We observe that entanglement assistance does not offer any advantage in previous scenarios. With realistic parameters we do not arrive at the parameter regime where the \ac{OEAR} shines (this region is reached when $n/b \ll 1$ and $\nu \gg 1$). The beneficial operating region can however be achieved with the \ac{mMIMO} \cite{lu2014overview} approach. In this operating regime transmitter and receiver are composed of multiple antennas. With fixed total power $n'= n/s$ of the antenna array one reduces the photon number per pulse and antenna by increasing the number $s$ of antennas per \ac{RCU} link and by assuming that antennas do not coordinate their transmission amongst different sender/receiver pairs. In all scenarios below we consider noise levels of a pure loss factory with size $K=100$, with grid distance $d=10m$. The cumulative capacity is calculated by multiplying capacity of interest (equations \eqref{eq:C_E}, \eqref{eq:C_J} and \eqref{eq:C_S}) with the number of subdivisions $s$.


\subsection{\ac{OFDM} factory with \ac{OFDM} type \ac{mMIMO}}

Our factories would not benefit from the \ac{mMIMO} with \ac{OFDM} communication scheme. Considering any capacity from above:
\begin{align}
    C^{mMIMO}(b,\tau,n,\nu,s) &= s \cdot C(b',\tau,n',\nu')
\end{align}
In the \ac{OFDM} type \ac{mMIMO}, the baud-rate $b'=B/(s \cdot K^2)$ is affected by factory size and the number of sub-links. The new power per pulse $n'/b'=\frac{n/s}{B/(s\cdot K^2)}$ is unaffected by the sub-link number, as is the new noise per pulse $\nu'=\nu_{b}/(s\cdot b')$. Further-more, the $s$ multiplication factor in $C^{mMIMO}$ and the $b'$ baud-rate cancel out. Thus the capacity achieved with \ac{OFDM} type \ac{mMIMO} is the same as the capacity achieved by \ac{OFDM} link without \ac{mMIMO}. The metrics relevant for the \ac{OEAR} and \ac{OJDR} performance are also unaffected.


\subsection{\ac{OFDM} factory with \ac{CDMA} type \ac{mMIMO}}

In contrast to \ac{OFDM} type \ac{mMIMO} the \ac{CDMA} type \ac{mMIMO} offers \ac{OEAR} scaling with \ac{mMIMO} array size $s$. Each link has a bandwidth $B^{OFDM}=B/K^2=10^8$. In the \ac{CDMA} case the baud-rates are not affected by the $s$ parameter. The power per pulse ratio $n'/b=n/(s\cdot B^{OFDM})$ is scaling favourably for the \ac{OEAR}, reaching the $n'/b < 1$ at $s\approx 10^4$. We assume antenna coupling between the sub-links is identical between all transmitter-receiver pairs. The noise is then calculated as
\begin{align}
    \nu' &= \frac{\nu_{TH} \cdot B^{OFDM} }{s} + \tau n \frac{s-1}{s}
\end{align}
\begin{figure}[H]
    \centering
    \input{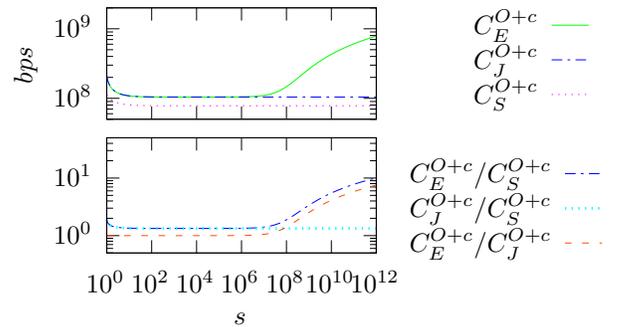}
    \caption{Logarithmic scaling of capacity with increasing numbers of the antenna array with \ac{mMIMO} approach in robot factory with hybrid \ac{CDMA} and \ac{OFDM} approach.}
    \label{fig:results_mMimo}
\end{figure}

\subsection{\ac{CDMA} factory with \ac{OFDM} type \ac{mMIMO}}
The case where the \ac{OFDM} type \ac{mMIMO} is used within the \ac{CDMA} factory, is no different to the same \ac{mMIMO} type inside of the \ac{OFDM} factory. In the same way all capacities are unchanged by the array size factor $s$.

\subsection{\ac{CDMA} factory with \ac{CDMA} type \ac{mMIMO}}

The \ac{CDMA} factory with \ac{CDMA} type \ac{mMIMO} benefits from high constant baud-rates $b=B$ as well as low power per pulse $n/b=n/(B \cdot s)$, which is also dependent on the \ac{mMIMO} array number $s$. In this case again for $s=1$ the ratio  $\tau n'/b \approx 10 \cdot 10^{-4}$ where \ac{QIP} already offers an advantage. With link subdivision, the ratio is improved further. 

\begin{figure}[H]
    \centering
    \input{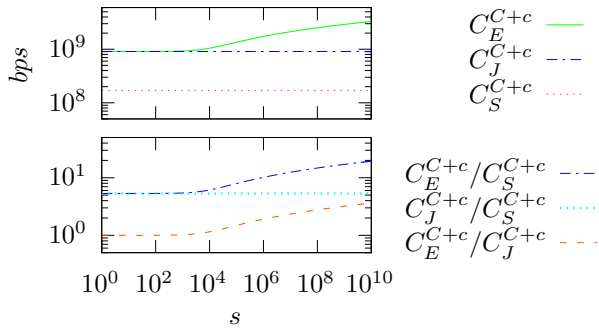}
    \caption{Logarithmic scaling of capacity with increasing numbers of the antenna array with \ac{mMIMO} approach in robot factory with hybrid \ac{CDMA} and \ac{OFDM} approach.}
    \label{fig:results_mMimo}
\end{figure}

Values of $s$ at around $10^{4}$ mark the ratio where $n'/b$ falls below $1$, thus at this point we start to observe the \ac{OEAR} advantage over \ac{OJDR} receiver. 





\section{Conclusions}

In this paper we have investigated communication scaling of large uncoordinated networks. We pointed out possible advantages in terms of capacities in the order of a factor $10$ over the state of the art. 

We modeled an increase of noise with the factory size in a simple but representative way. Two noise scaling models were considered, one which slightly over- and another which slightly underestimated the interference power.

It is often argued that noise decreases quantum advantages. However we observe only a weak dependence of the \ac{QIP} advantage on the accumulated interference power.

Throughout, our analysis shows that \ac{CDMA} is to be preferred over \ac{OFDM}: In all cases \ac{CDMA} kept its improvement scaling with a factor of $K^2$ in comparison to the equivalent \ac{OFDM} technology. This observation holds true when one considers baud-rates in between $10^{11}Hz$ and $10^{12}Hz$ which are feasible today. The \ac{OEAR} theoretically offers the most beneficial capacity scaling in all cases, but it becomes visible only when an uncoordinated \ac{mMIMO} approach is introduced for each individual link. 

With this work we have outlined possible theoretical alternative solution to the increasing number and density of communicating devices in a factory-like setting.


\section*{Acknowledgment}
Funding from the Federal Ministry of Education and Research of Germany /BMBF) via grants 16KISQ039, 16KIS1598K, and "Souver\"an. Digital. Vernetzt." joint project 6G-life, grant number: 16KISK002, DFG Emmy-Noether program under grant number NO 1129/2-1 (JN) and the support of the Munich Center for Quantum Science and Technology (MCQST) are acknowledged.

\printbibliography
\end{document}